\documentclass[prd,a4paper,twocolumn,superscriptaddress]{revtex4}
\usepackage{graphicx}
 
\newcommand{\be}{\begin{equation}}
\newcommand{\ee}{\end{equation}}
\newcommand\lsim{\mathrel{\rlap{\lower4pt\hbox{\hskip1pt$\sim$}}
    \raise1pt\hbox{$<$}}}
\newcommand\gsim{\mathrel{\rlap{\lower4pt\hbox{\hskip1pt$\sim$}}
    \raise1pt\hbox{$>$}}}
\newcommand\esim{\mathrel{\rlap{\raise2pt\hbox{\hskip0pt$\sim$}}
    \lower1pt\hbox{$-$}}}

\begin{document}

\title{Linearized Bekenstein Varying Alpha Models}

\author{P.P. Avelino}
\email[Electronic address: ]{ppavelin@fc.up.pt}
\affiliation{Centro de F\'{\i}sica do Porto, Rua do Campo Alegre 687,
4169-007, Porto, Portugal}
\affiliation{Departamento de F\'{\i}sica da Faculdade de
Ci\^encias da Universidade do Porto, Rua do Campo Alegre 687,
4169-007, Porto, Portugal}
\author{C.J.A.P. Martins}
\email[Electronic address: ]{C.J.A.P.Martins@damtp.cam.ac.uk}
\affiliation{Centro de F\'{\i}sica do Porto, 
Rua do Campo Alegre 687, 4169-007, Porto, Portugal}
\affiliation{Department of Applied Mathematics and Theoretical
Physics, Centre for Mathematical Sciences,\\ University of
Cambridge, Wilberforce Road, Cambridge CB3 0WA, United Kingdom}
\author{J.C.R.E. Oliveira}
\email[Electronic address: ]{jeolivei@fc.up.pt}
\affiliation{Centro de F\'{\i}sica do Porto, Rua do Campo Alegre 687,
4169-007, Porto, Portugal}
\affiliation{Departamento de F\'{\i}sica da Faculdade de
Ci\^encias da Universidade do Porto, Rua do Campo Alegre 687,
4169-007, Porto, Portugal}
 
\date{16 February 2004} 
\begin{abstract} 
We study the simplest class of Bekenstein-type, varying $\alpha$ models,
in which the two available free functions (potential and gauge kinetic
function) are Taylor-expanded up to linear order. Any realistic model of
this type reduces to a model in this class for a certain time interval
around the present day. Nevertheless, we show that no such model 
is consistent with all existing observational results. We discuss possible
implications of these findings, and in particular clarify the ambiguous
statement (often found in the literature) that ``the Webb results are
inconsistent with Oklo''.
\end{abstract} 
\preprint{tba} 
\maketitle 
 
\section{Introduction} 
In theories with additional spacetime dimensions \cite{Polchinski}
there are typically many light or massless degrees of freedom, which can give
rise to a number of observable cosmological consequences. Noteworthy among
these are  variations of the fundamental couplings \cite{Essay,Uzan,VFC} 
(with the ensuing violations of the Equivalence Principle \cite{Will,Damour2})
and contributions to the energy density budget of the Universe. 
 
In recent years there has been a growing body of evidence for the presence 
of these two effects. Type Ia supernovae \cite{Tonry},
the Cosmic Microwave Background (CMB) \cite{Bennett}
and lensing data \cite{Bernardeau} are
all consistent with the existence of the so-called 
dark energy component, whose gravitational behaviour is very similar to that 
of a cosmological constant, and which indeed appears to have become the 
dominant component in the energy budget of the Universe 
at a redshift $z\sim1$. 
 
On the other hand there is some (somewhat more controversial) evidence for 
the spacetime variation of the fine-structure constant $\alpha$, coming 
from both quasar absorption systems
(at redshifts $z\sim1-3$, \cite{Webb,Murphy,Petit})
and the Oklo natural nuclear reactor ($z\sim0.1$, \cite{Fujii}).
There is also a further claim of a varying proton to electron mass ratio,
also at $z\sim3$ \cite{Ivanchik}. While it is conceivable 
that hidden systematic effects are still contaminating some of these 
measurements, an unprecedented effort is being made by a number of independent 
groups and using a range of techniques, to search for such variations at 
various key cosmological epochs, which should soon clarify the situation.
There is also a range of other constraints, either
local \cite{Marion} (from atomic clocks)
or at low \cite{Olive} (from Rhenium decay in meteorites) or high redshift \cite{Avelino,Martins,Rocha} (from the CMB and BBN),
with much stringent ones forthcoming \cite{wmap1,wmap2}.
 
It goes without saying that from the point of view of a fundamental theory 
there is more than ample freedom to allow the dark energy and the varying
couplings to be due to different degrees of freedom in the theory, and even
to emerge through different physical mechanisms. 
Nevertheless, it is useful to study the simplest case in which the two have a 
common origin, as this will in principle have the fewest free parameters 
and can therefore be better constrained. 
 
In what follows we will discuss a very simple toy model for this which, 
although arguably oversimplified from the particle physics point of view, has 
the advantage of having a minimal number of free parameters. The basic 
idea is to consider Bekenstein-type models \cite{Bekenstein},
and reduce the freedom in the two free functions (the potential $V(\phi)$
and the gauge kinetic function $B_F(\phi)$) by Taylor-expanding them around
the present day, and retaining only terms up to linear order. In fact, we will 
see that its free parameters are in some sense too few, so that the model is 
very tightly constrained by existing observations, and indeed ruled out if
all of them are correct. Still the model can be useful in providing some
guidance for the likely requirements of successful, fundamental theory
inspired models. Other interesting analyses of this class of models can be
found in \cite{Sandvik,OlivePos,Anchordoqui,Parkinson,Copeland,Nunes}. 

We will start in Sect. II with a brief overview of the Bekensein-type models.
We then discuss in more detail the linearized regime that interests us
(Sect. III) and discuss it in the context of existing observational data
(Sect. IV). Finally Sect. V summarizes our results and briefly discusses
further prospects.
Throughout this paper we shall use fundamental units with $\hbar=c=G=1$.
 
\section{Overview of Bekenstein models} 
 
The variation of the fine-structure constant in
Bekenstein-type theories \cite{Bekenstein} is 
due to the coupling of a scalar field $\phi$ to the electromagnetic field 
tensor $F_{\mu\nu}$, through a term of the generic form 
\be 
S_{em}=\int d^4x\sqrt{-g}\left[-\frac{1}{4}B_F\left(\frac{\phi}{m}\right) 
F_{\mu\nu}F^{\mu\nu}\right]\,, 
\label{coupling} 
\ee 
where $m\sim m_{Pl}=1$ and $B_F$, known as the gauge kinetic function, is the 
effective dielectric permittivity. This should be explicitly specified in 
a fundamental theory, but can be phenomenologically taken to be a free 
function. The fine-structure constant is then given by 
\be 
\alpha(\phi)=\frac{\alpha_0}{B_F(\phi/m)}\,, 
\label{gkfalpha} 
\ee 
and at the present day one has $B_F(\phi_0/m)=1$. 
 
The inclusion of an interaction term such as $B_F(\phi)F^2$ that is 
non-renormalizable in 4D requires, at the quantum level, the existence of an 
ultra-violet momentum cutoff. Any particle physics motivated 
choice of this cutoff will destabilize the quintessence potential,
since it will yield a mass term much larger than the quintessence one
(recall that typically $m_q\sim H_0$). This 
is therefore a further fine-tuning problem, akin to the cosmological 
constant one. In our phenomenological approach (and in common with all
previous work on these models) we will ignore this problem, 
assuming that any mechanism that solves the cosmological constant problem  
will also solve this one. (Such a mechanism must exist if the dark energy
of the universe is indeed quintessence-like.)
 
Assuming that the cosmological change in the scalar field $\phi$ is small (at 
least in recent epochs), one can expand all couplings around their present-day 
values, in particular 
\be 
B_F\left(\frac{\phi}{m}\right)=1+\zeta_F\frac{\phi-\phi_0}{m}+\frac{1}{2}\xi_F 
\left(\frac{\phi-\phi_0}{m}\right)^2+\ldots\,, 
\label{expandbf} 
\ee 
corresponding to a variation of $\alpha$ (again relative the present day) 
\be 
-\frac{\Delta\alpha}{\alpha}=\zeta_F \frac{\phi-\phi_0}{m}+\frac{1}{2} 
(\xi_F-2\zeta_F)\left(\frac{\phi-\phi_0}{m}\right)^2+\ldots\,. 
\label{generalalpha} 
\ee 
Given that the classical predictions of the model will be independent of the 
particular choice for the mass $m$, we shall take $m=m_{Pl}=1$ throughout.
In addition to the variation of the fine-structure constant, this coupling 
is responsible for an effective non-universality of the gravitational force, 
which through Equivalence Principle tests \cite{Will,Damour2}
leads to the constraint 
\be 
\left| \zeta_F \right|<5\times 10^{-4}\,; 
\label{wepconstraint} 
\ee 
note, for example, that in Bekenstein's original theory one has $\zeta_F=-2$. 

The evolution of the scalar field is then typically of the form 
\be 
{\ddot \phi}+3H{\dot \phi}+\frac{dV}{d\phi}=-\zeta_m \rho_m\,, 
\label{genevol} 
\ee 
where $\rho_m$ is the matter density of the universe. Given a complete particle 
theory, $\zeta_F$ will be specified and it should be possible to calculate the 
coupling of the scalar field to matter, $\zeta_m$. Typically the scalar field 
will only evolve significantly during the matter era---a result that is
well known from the study of scalar-tensor theories \cite{Damour,Santiago}.
A non-zero $\zeta_F$ is 
in principle sufficient to ensure a cosmological variation of $\phi$, driven 
by the electromagnetic part of the baryon mass density, and hence a variation 
of $\alpha$. However, the resulting change will typically be small if this is 
the only source. For example, in the original Bekenstein model, one can only
fit the Webb results at the cost of having a huge violation of the weak equivalence principle; conversely if one wants to satisfy these constraints
the typical allowed variation is only $\Delta\alpha/\alpha\sim10^{-10}$. 
This constraint can only be evaded by saying that $\phi$ couples only 
(or predominantly) do the dark matter \cite{Sandvik}. 
 
In addition to the equivalence principle constraint, there are a number of 
bounds or detections which restrict the cosmological evolution of $\alpha$, 
as already discussed above. Moreover, if one assumes that $\phi$ is also 
providing the dark energy its evolution will be further constrained through 
its present contribution for the energy budget and the evolution of its 
equation of state \cite{Steen,Melchiorri,Jimenez}. 
 
\section{Analysis of the Linearized Case} 
 
Let us consider the class of models of a neutral scalar field coupled  
to the electromagnetic field with
\begin{equation} 
{\cal L} = {\cal L}_\phi + {\cal L}_{\phi F} + {\cal L}_{\rm other}\, , 
\end{equation} 
where
\begin{equation} 
{\cal L}_\phi= \frac{1}{2}\partial^\mu \phi \partial_\mu \phi - V(\phi)\, , 
\end{equation} 
\begin{equation} 
{\cal L}_{\phi F}= -\frac{\alpha_0}{4 \alpha} F_{\mu \nu} F^{\mu \nu}\, , 
\end{equation} 
and ${\cal L}_{\rm other}$ is the Lagrangian density of the other fields. 
We will make the simplifying assumption that both $V(\phi)$  
and $\alpha$ are linear functions of $\phi$, namely 
\begin{equation} 
V(\phi) = V(\phi_0) + \frac{dV}{d\phi} \left(\phi-\phi_0\right)\, , 
\end{equation} 
and 
\begin{equation} 
\alpha = \alpha_0 + \frac{d\alpha}{d\phi} \left(\phi-\phi_0\right)\, , 
\end{equation} 
with both ${dV}/{d\phi}$ and ${d\alpha}/{d\phi}$ assumed to be  
constants. One then has 
\be 
\frac{\Delta\alpha}{\alpha}=\left(B_F^{-1}-1\right)=\frac{1}{\alpha_0} 
\frac{d\phi}{d\alpha}(\phi-\phi_0)\,. 
\label{gkfspec} 
\ee 
 
We will further assume that $dV/d\phi < 0$.  
Assuming that the interpretation of the Webb \textit{et al.}
results as evidence for a variation of the fine structure constant  
is correct this implies a smaller value of $\alpha$ in the past and  
consequently ${d\alpha}/{d\phi} > 0$, though if the claimed Oklo detection is 
also true there must be oscillations. Let us first comment on 
these assumptions. In any model both $V(\phi)$ and $\alpha(\phi)$ can be  
taken as linear functions of $\phi$ for some limited period of time around
today. For how long that assumption holds is of course model dependent.
In this paper we will be considering a particular class 
of models for which this assumption is valid for a considerable time, possibly
even all the way from the epoch of nucleosynthesis up to the present time.  
In other words, we are effectively testing the validity of this assumption
for the class of Bekenstein models, assuming the validity of the claimed
low-redshift detections of a varying fine-structure constant.

In a spatially flat Friedman-Robertson-Walker universe the equations of  
motion are given approximately by 
\begin{equation} 
H^2=H_0^2\left(\Omega_{m0} a^{-3}+\Omega_{r0} a^{-4}+\Omega_{\Lambda0} +  
\Omega_\phi \right)\, , 
\end{equation} 
\begin{equation} 
\frac{\ddot a}{a}=-H_0^2\left[\frac{\Omega_{m0}}{2} a^{-3} 
+\Omega_{r0} a^{-4}-\Omega_{\Lambda0} +  
\frac{\Omega_\phi}{2} (1+3w_\phi)\right]\, , 
\end{equation} 
where  
\begin{equation} 
\Omega_\phi=\frac{8\pi ({\dot \phi}^2/2+V(\phi))}{3 H_0^2}\, , 
\end{equation} 
and  
\begin{equation} 
\omega_\phi=\frac{{\dot \phi}^2/2-V(\phi)}{{\dot \phi}^2/2+V(\phi)}\, .  
\end{equation} 
Since any variation of the fine structure constant from the epoch of  
nucleosynthesis onwards is expected to be very small \cite{Avelino,Rocha}
we have also neglected the minor contribution that such a variation has in
the evolution of the baryon density (included in $\Omega_{m0}$). 
The equation of motion for the field $\phi$ is given by 
\begin{equation} 
\label{phieq} 
{\ddot \phi}+3H{\dot \phi}=-\frac{dV}{d \phi}-\frac{\alpha_0}{4 \alpha^2} 
\frac{d \alpha}{d \phi} 
F_{\mu \nu} F^{\mu \nu} \, . 
\end{equation} 
It is crucial to discuss the relative importance of the last two terms.  
We shall assume that the main contribution to the last term comes from  
baryons. Given that $F^{\mu \nu} F_{\mu \nu}= 2 (B^2-E^2)<0$ the last  
two terms in equation (\ref{phieq}) have opposite signs. It has been shown 
that the time variations  
of the fine structure constant induced by the last term are two small to  
ever be observed (if Equivalence Principle constraints are to be
obeyed \cite{OlivePos}). However, this does not happen with the first term. 
Hence, in order to have interesting variations of $\alpha$ the first term  
needs to dominate at recent times which will happen if 
\begin{equation} 
\left(\frac{dV}{d \alpha}\alpha\right)_0 \gg   
\left(\frac{1}{4}F_{\mu \nu} F^{\mu \nu}\right)_0 \sim 10^{-4} \rho_{c0}\, , 
\end{equation} 
where $\rho_{c0}$ is the present-day critical density.
Note that the last term can be neglected at the present time but it  
becomes important at early times since the main contribution to this term  
comes from baryons whose energy density varies as $a^{-3}$.  
In summary, in this context, a particular model is fully specified by the  
parameters ${\dot \phi}_0$, $V_0$, $dV/d\phi$ and $d \alpha / d \phi$, 
in addition to the nominal cosmological parameters.  
 
Let us start with some order of magnitude constraints on the value of  
$\delta \phi \equiv \phi(z=0)-\phi(z=1)$, with $z\sim 1$ being singled out
as the approximate redshift at which the matter domination epoch ends
(do not confuse this particular $\delta\phi$ with the generic $\Delta\phi$
that has already been introduced above). If our theory is to produce  
interesting variations of the fine structure constant with redshift  
the first term in the right hand side of (\ref{phieq}) needs to be the  
dominant one. Hence we have  
\begin{equation} 
\delta \phi H^2 \sim \frac{\delta V}{\delta \phi} \lsim  
\frac{\rho_{c0}}{\delta \phi}\, . 
\end{equation} 
which implies that $\delta \phi \lsim 10^{-1}$. On the other  
hand if we want to have a variation in the fine structure constant of  
$\delta \alpha / \alpha \sim 10^{-6}$ without violating the equivalence  
principle one needs  
\begin{equation} 
\frac{\delta \alpha}{\alpha} = \frac{\delta \phi}{\alpha} 
\frac{d \alpha}{d\phi} \sim 10^{-6}\, . 
\end{equation} 
Given that the equivalence principle tests give the constraint 
\begin{equation} 
\frac{d \alpha}{d\phi} \lsim 10^{-5}\, , 
\end{equation} 
one has necessarily $\delta \phi \gsim 10^{-3}$. These two constraints give  
a limit on $\delta \phi$ of 
\begin{equation} 
10^{-3} \lsim \delta \phi \lsim 10^{-1}\, , 
\end{equation} 
which will be further tightened by future tests of the equivalence  
principle. A related limit on $dV/d\phi$ gives: 
\begin{equation} 
-H_0^2 \lsim \frac{dV}{d\phi} \lsim -10^{-2} H_0^2\, , 
\end{equation} 
In order for the lower limit on $\delta \phi$ to be satisfied  
one needs $w_\phi(z=0)-1 \gsim 10^{-4}$. It is also straightforward  
to verify that $w \to -1$ very rapidly as we move backwards in time. 
 
\begin{figure} 
\includegraphics[width=3.5in,keepaspectratio]{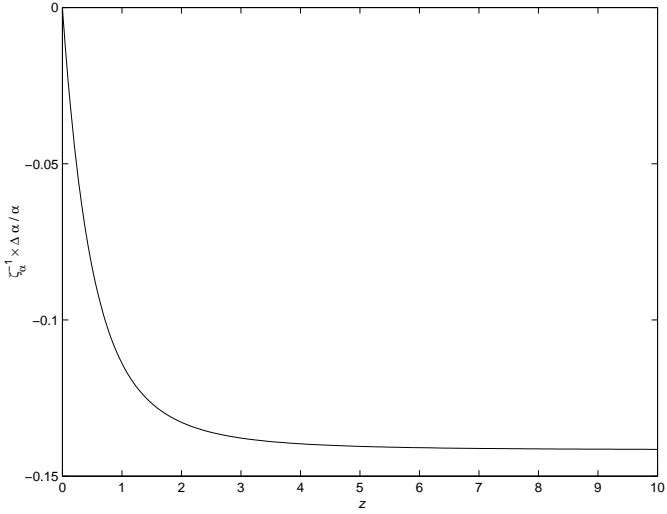} 
\caption{\label{fig1}The evolution of the value of  
$\Delta \alpha / \alpha$ as a function of redshift, in re-scaled units 
with $\zeta_\alpha \equiv \frac{1}{\alpha H_0^2}\frac{dV}{d\alpha}
=-\zeta_F \frac{1}{H_0^2}\frac{dV}{d\phi}$.} 
\end{figure} 

\begin{figure} 
\includegraphics[width=3.5in,keepaspectratio]{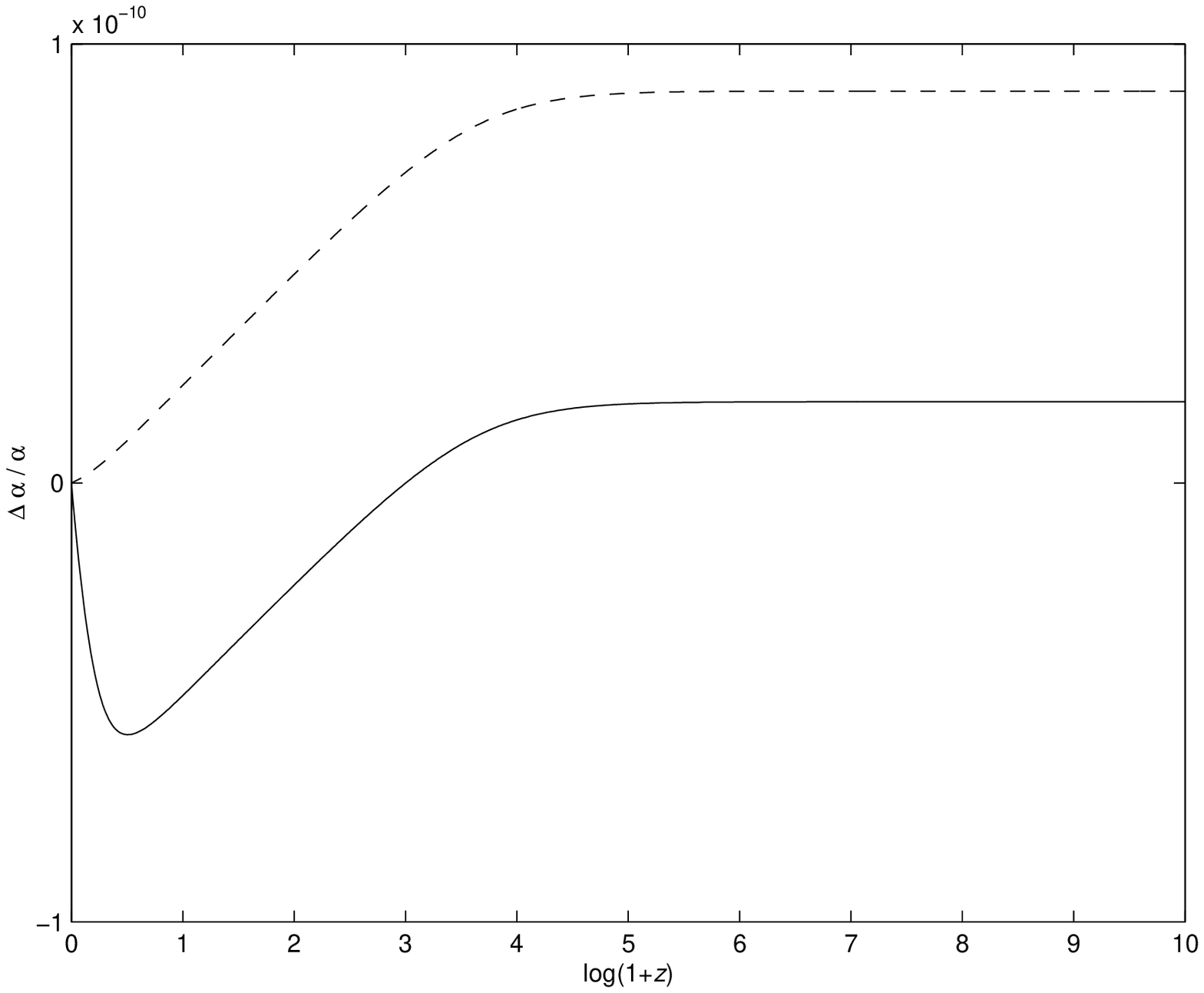} 
\includegraphics[width=3.5in,keepaspectratio]{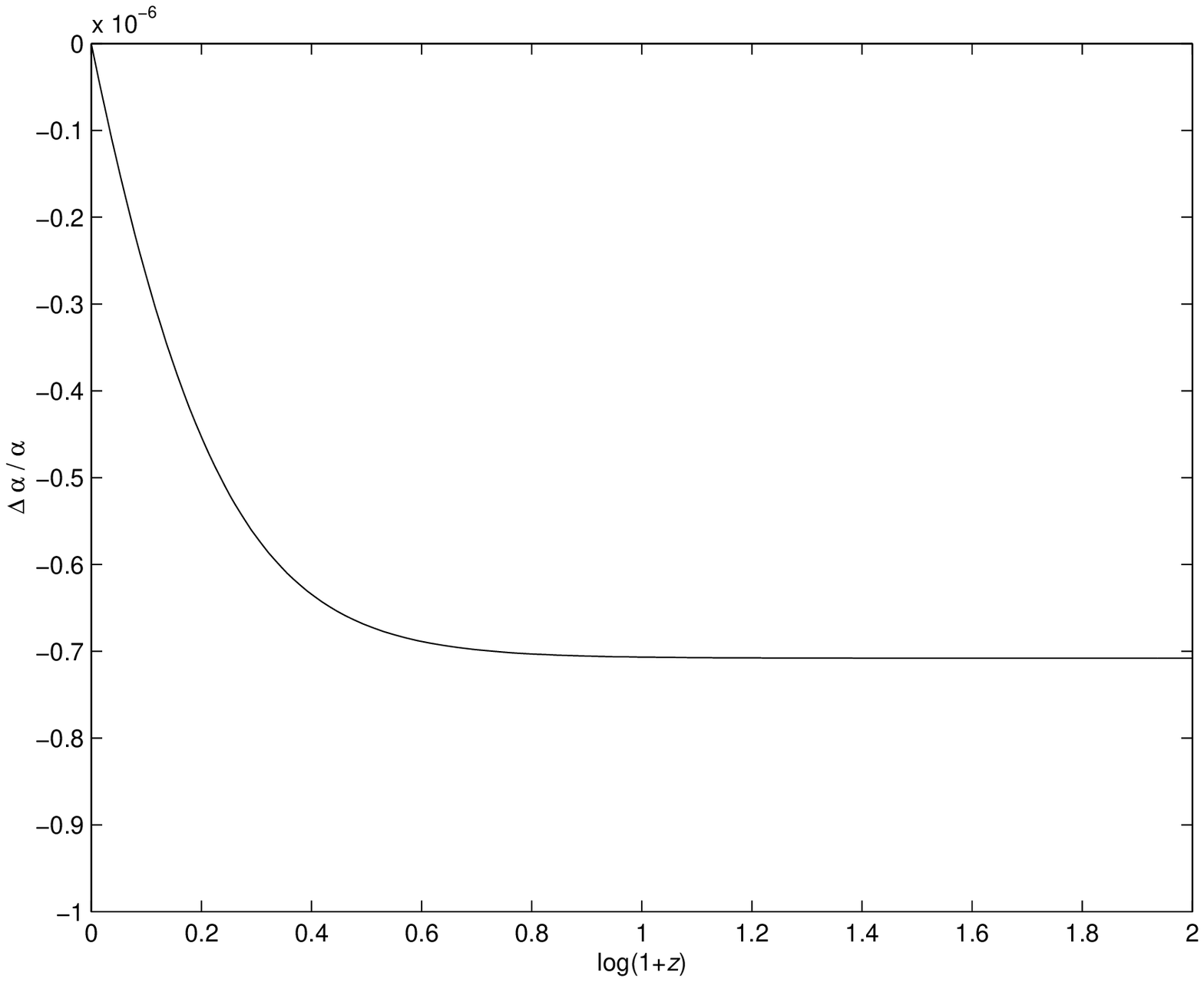} 
\caption{\label{fig2} {\bf (Top Panel):} The evolution of the value of  
$\Delta \alpha / \alpha$ as a function of redshift for 
$dV/d\phi=-10^{-6} H_0^2$ (solid line) and $dV/d\phi=0$ (dashed line) 
with $\zeta_F=-5 \times 10^{-4}$ and $\Theta=10^{-8} H_0^2$.  
The electromagnetic 
term gives a negligible contribution to the variation of 
$\alpha$ if $|dV/d\phi|$ is large enough favouring a larger value of 
$\alpha$ in the past. {\bf (Bottom Panel):} The evolution of the value of  
$\Delta \alpha / \alpha$ as a function of redshift for 
$dV/d\phi=-10^{-2} H_0^2$. We clearly see that no cosmological 
significant variation of $\alpha$ exists beyond $z=10$.} 
\end{figure}

\begin{figure} 
\includegraphics[width=3.5in,keepaspectratio]{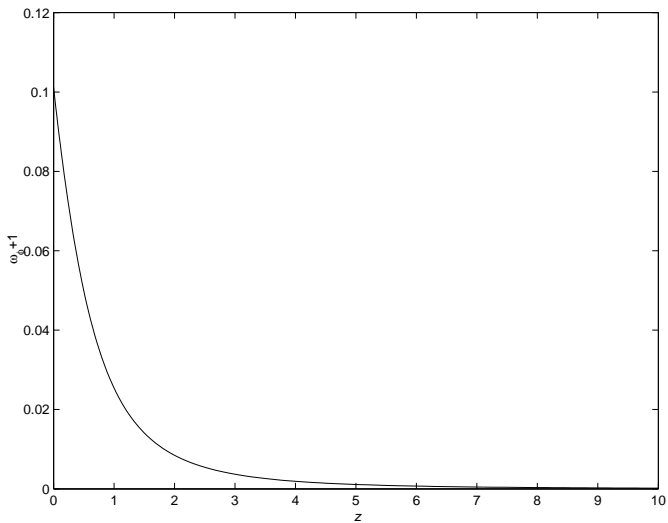} 
\caption{\label{fig3}The evolution of the value of  
$w_\phi+1$ with redshift, $z$, for $dV/d\phi=-0.35$ assuming  
$\Omega_{\phi0} \sim 0.7$ and $w^0_\phi \sim -1$. Note that $w_\phi + 1$ 
evolves very rapidly toward zero when one moves backward in time.} 
\end{figure} 

\begin{figure} 
\includegraphics[width=3.5in,keepaspectratio]{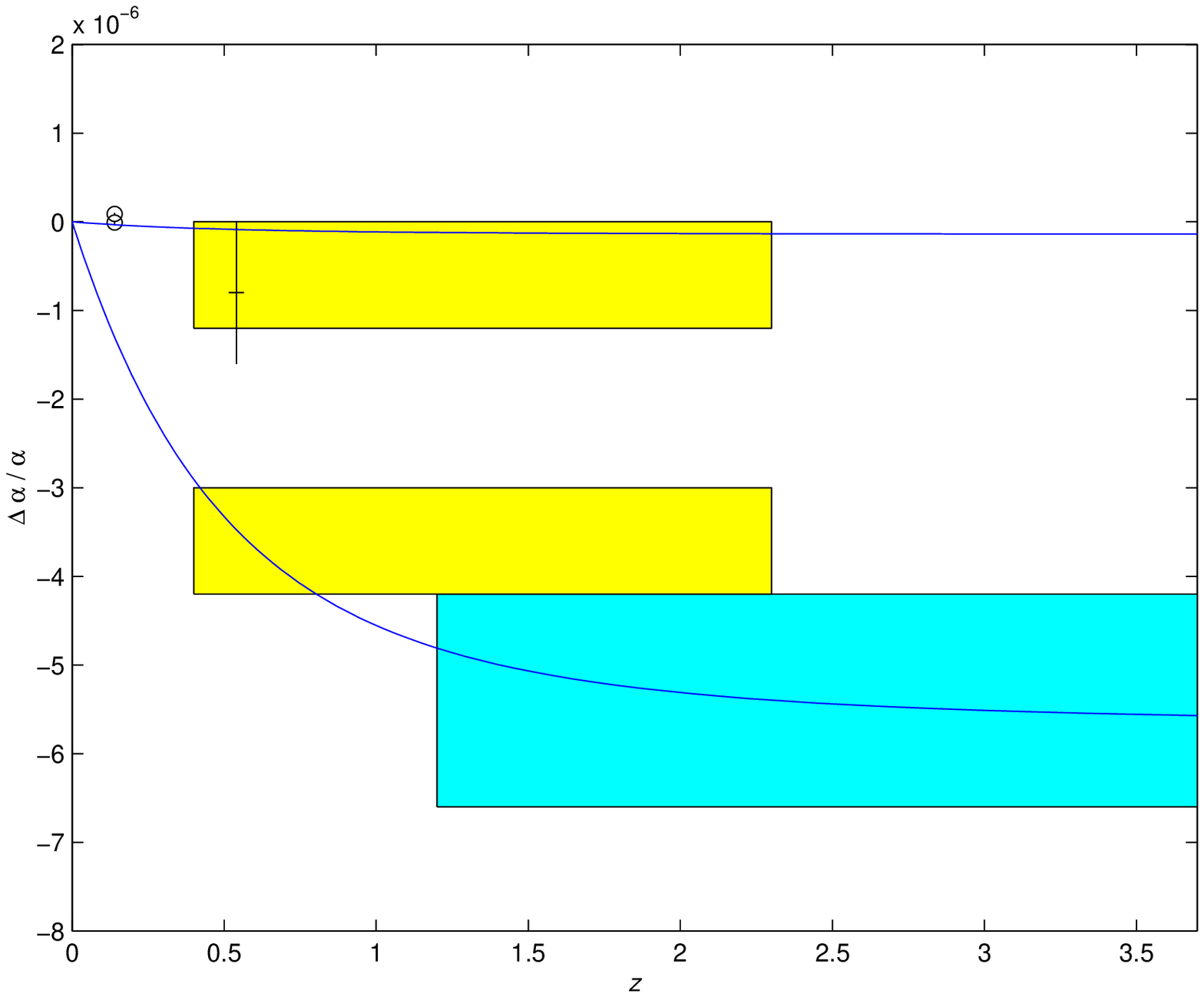} 
\caption{\label{fig4}Comparison of two typical examples of this
class of models with observational data: Oklo (\cite{Fujii}, circles),
Rhenium decay from meteorites (\cite{Olive}, vertical bar)
and quasar data (Murphy \textit{et al.} \cite{Murphy} dark shaded box,
Chand \textit{et al.} \cite{Petit} light shaded box). Either
this class of models is not valid up to redshifts about $z\sim3$, or some
of the data is strongly affected by systematics.
See main text for further discussion.} 
\end{figure} 
 
Before we investigate the possible role of the scalar field responsible for  
the variation of the fine structure constant as a quintessence candidate it  
is instructive to study the evolution of the scalar field assuming that its  
contribution to the dynamics of the universe is subdominant. From the
equations above it is easy to show that 
\begin{equation} 
f(a)\frac{d^2 \phi}{da^2}+g(a)\frac{d \phi}{da}=\frac{1}{H_0^2} 
\left(-\frac{dV}{d \phi}-\frac{\Theta}{a^3}\right) \, , 
\end{equation} 
where 
\begin{equation} 
f(a)=\Omega_{m0} a^{-1}+\Omega_{\Lambda0} a^{2}+\Omega_{r0} a^{-2}\, , 
\end{equation} 
\begin{equation} 
g(a)=\frac{5}{2}\Omega_{m0} a^{-2}+4\Omega_{\Lambda0} a+2\Omega_{r0} a^{-3}\,, 
\end{equation} 
and the last term has been expressed as a function of the behaviour of the 
matter density (with the constant $\Theta$ absorbing the additional parameters). 
Assuming that $dV/d\phi$ is a constant one can find the following asymptotic
solutions in the radiation, matter and $\Lambda$-dominated epochs 
\begin{equation} 
\phi_r=A_r+B_r a^{-1}+\frac{1}{\Omega_{r0} H_0^2}\left(-\frac{1}{20} 
\frac{dV}{d \phi} a^4-\frac{\Theta}{6}a\right) \, , 
\label{solrad} 
\end{equation} 
\begin{equation} 
\phi_m=A_m+B_m a^{-3/2}+\frac{1}{\Omega_{m0} H_0^2}\left(-\frac{2}{27} 
\frac{dV}{d \phi} a^3-\frac{2\Theta}{3}\ln{a}\right) \, , 
\label{solmat} 
\end{equation} 
\begin{equation} 
\phi_\Lambda=A_\Lambda+B_\Lambda a^{-3}+\frac{1}{3\Omega_{\Lambda0} H_0^2}\left[
-\frac{dV}{d \phi}\ln{a}+\Theta a^{-3}
\left(\ln{a}+\frac{1}{3}\right)\right] \, . 
\label{sollam} 
\end{equation} 
  
If we neglect the decaying  
mode it is possible to match the solutions deep in the matter era  
with the solution deep in the radiation era   
era in such a way that both $\phi$ and $d\phi/da$ are continuous  
functions of the redshift. We can then find  
$A_\Lambda$, $B_\Lambda$, $A_m$ as
\begin{equation} 
A_\Lambda=A_m-\frac{1}{27 H_0^2\Omega_{\Lambda0}} \frac{dV}{d \phi}  
\left[1-6\ln\frac{\Omega_{m0}}{\Omega_{\Lambda0}}+\frac{\Theta}{3}
\left(2+\ln\frac{\Omega_{m0}}{\Omega_{\Lambda0}}\right) \right] \, , 
\end{equation} 
\begin{equation} 
B_\Lambda=-\frac{1}{27 H_0^2\Omega_{m0}} \frac{dV}{d \phi}  
\left(\frac{\Omega_{m0}}{\Omega_{\Lambda0}}\right)^2 
\left(1+3\ln\frac{\Omega_{m0}}{\Omega_{\Lambda0}}+\frac{\Theta}{3}
\frac{\Omega_{m0}}{\Omega_{\Lambda0}}\right) \, , 
\end{equation} 

As expected \cite{Damour,Santiago,Sandvik}, we find that the evolution of
$\phi$ is negligible during the radiation dominated epoch but significant
during the matter one, and that the onset of cosmological constant domination 
damps this evolution. On the other hand, the cosmological evolution in the
case where the dark energy of the universe is provided by $\phi$ itself
rather than by a cosmological constant can be approximately inferred from
the above analysis, since it is observationally known that the gravitational
behaviour of the former must be very close to the latter. 

\section{Discussion}

In Fig. \ref{fig1} we plot the evolution of the fine structure constant as 
a function of redshift. Given that ${\dot \phi}_0$ is proportional to 
$dV/d\phi$ and the dynamics of the universe near the present time is 
constrained to be very close to that of a universe with 
$\Omega_\phi^0 \sim 0.7$ and  $\omega_\phi \sim -1$, the shape of the 
evolution curve of $\Delta \alpha / \alpha$ as a function of redshift, $z$, 
is unambiguously predicted by our model up to a normalization factor
\begin{equation}
\zeta_\alpha\equiv- \frac{dV}{d\phi} \frac{\zeta_F}{H_0^2} \,.
\label{norm}
\end{equation}
We see that for a value of 
$\zeta_\alpha\sim -10^{-6}$ (take for example $\zeta_F = - 10^{-4}$ and
$dV/d\phi=-10^{-2} H_0^2$ one has interesting 
variations of $\alpha$ which we would expect to be able to detect some 
time in the near future. However, note that the evolution of $\alpha$
with redshift is still quite significant at very low redshifts, which
indicates that it is not possible to reconcile the Oklo \cite{Fujii} or
meteorite \cite{Olive} results (at $z=0.14$ and $z=0.45$ respectively)
with the Webb/Murphy results \cite{Webb,Murphy} in the context of our model.
If we take the Oklo and meteorite limits seriously the maximum variation of
$\alpha$ that is allowed in this class of models by a redshift $z \sim 2$ say
is about $10^{-7}$, clearly below the Webb/Murphy results, though perhaps
compatible with Chand \textit{et al.} \cite{Petit}. More on this below.

Examples of two particular models are displayed in Fig. \ref{fig2}.
In the top panel we plot the evolution of the value of  
$\Delta \alpha / \alpha$ as a function of redshift for 
$dV/d\phi=-10^{-6} H_0^2$ (solid line) and $dV/d\phi=0$ (dashed line) 
with $\zeta_F=-5 \times 10^{-4}$ and $\Theta=10^{-8} H_0^2$.  
For a nearly flat  
scalar field potential the electromagnetic term in equation (\ref{phieq}) 
is the main source of a variation of the fine structure constant, 
favouring a larger value of $\alpha$ in the past. However, in that case the 
variations are too small to be of any cosmological significance. 
In the bottom panel we plot the evolution of the value of  
$\Delta \alpha / \alpha$ as a function of redshift for 
$dV/d\phi=-10^{-2} H_0^2$ showing that no cosmological 
significant variation of $\alpha$ beyond $z=10$.

Given a value of $H_0$, the parameters $V_0$ and ${\dot \phi}_0$ 
determine in a unique way $w_{\phi0}$ and $\Omega_{\phi0}$. Hence, the 
equation of state $w_\phi(z)$ is a function of $dV/d\phi$, $\Omega_{\phi0}$ 
and $\omega_{\phi0}$ only and will evolve very rapidly towards a cosmological  
constant with $\omega_\phi \to -1$ when we move backwards in time.  
This is clearly shown in  Fig. \ref{fig3} (again, note that $w_{\phi0}$ and 
$\Omega_{\phi0}$ are constrained to be very close to $-1$ and $0.7$
respectively).
 
Let us now go back to the issue of the comparison with observational data,
which is plotted in Fig. \ref{fig4} against two typical models. Note that
the Oklo \cite{Fujii} and meteorite \cite{Olive} data are plotted as points,
since they apply to specific redshifts. However, for the quasar absorption
data \cite{Murphy,Petit} we've chosen to plot them as bands, rather
than plotting the individual
observational points (or some binning thereof). This choice is partially
motivated by the fact that the error bars are still much larger than
those from Oklo and the meteorites, and also because the quasar `distilled'
results are often quoted as a single number that is supposed to apply to
a range of redshifts. Note however that this practice can be misleading.
For example, a number of authors quote the latest results by Webb and
collaborators \cite{Murphy} as
\begin{equation}
\frac{\Delta\alpha}{\alpha}=(-0.54\pm0.12)\times 10^{-5}\,,\quad 0.2<z<3.7\,,
\label{mmurphy}
\end{equation}
ignoring the fact that even though the sample spans all that redshift range,
the data only prefers a value of $\alpha$ different from today's beyond
redshift $z\sim1$---hence our choice for the horizontal range of the (dark
grey) band. For the Chand {\textit et al.} results (light grey bands) we have
included their two possible results,
\begin{equation}
\frac{\Delta\alpha}{\alpha}=(-0.06\pm0.06)\times 10^{-5}\,,\quad 0.4<z<2.3\,,
\label{chand1}
\end{equation}
assuming terrestrial isotopic abundances (case 1), or
\begin{equation}
\frac{\Delta\alpha}{\alpha}=(-0.36\pm0.06)\times 10^{-5}\,,\quad 0.4<z<2.3\,,
\label{chand2}
\end{equation}
assuming low-metalicity isotopic abundances (case 2). The true result is
expected to lie somewhere between the two.

A number of interesting points can be inferred from the figure. As a first
remark, let us point out that even though the Webb/Murphy and the
Chand {\textit et al.} results seem to be statistically inconsistent, 
the discrepancy
may be much smaller than one could guess by simply comparing (\ref{mmurphy})
and (\ref{chand1}). Nevertheless, the question remains as to whether there
are hidden systematics contaminating one or both of the data or the analysis
pipelines.

We have already pointed out that the distinguishing feature of this
class of models is that significant variations of $\alpha$ occur relatively
near the present epoch. (As a side remark we note that in this context this
justifies the commonly used assumption of a uniform value of $\alpha$ 
throughout the last scattering epoch when constraining variation of 
the fine structure constant with CMB
observations \cite{Avelino,Martins,Rocha,wmap1,wmap2}.) However, this late
variation has dramatic consequences. Roughly speaking, depending on the
model parameters one can divide models in this class into two different types:
they can either be consistent with Oklo$+$meteorites$+$(\ref{chand1}) but
be inconsistent with (\ref{mmurphy}), or else be consistent with
(\ref{mmurphy})$+$(\ref{chand2}), but inconsistent with Oklo$+$meteorites.

In other words, if we assume that our linearized class of models holds true
up to at least redshift $z\sim3$ or so, then the Webb/Murphy results are
indeed inconsistent with Oklo/meteorites. 
We note that a number of authors have in the past made the
unqualified statement that ``the Webb results are inconsistent with Oklo''.
Such a statement is not meaningful {\textit per se}, since any such comparison
is necessarily model-dependent: one needs to specify a timescale as well as
a model for the redshift evolution of $\alpha$ (see the discussion in
\cite{Essay}). Indeed one can build models where the two can be made
compatible---an example is \cite{Parkinson}.
Having said that, here we do find that the two are inconsistent
for the models we considered.

This therefore calls for improvements on the existing observational results.
Following the controversy generated by the quasar data results, at least five
(to our knowledge) independent groups are currently working on the subject,
using a variety of different methods, so there is hope that the situation
will be clarified soon. No similar interest exists for the Oklo or meteorite
data, though independent confirmation of both of these results would be much
welcome since as we have seen they are quite more constraining, particularly
for the class of models that we have discussed. We note that measurements
of $\alpha$ using quasar data are, notwithstanding the possible sources
of observational systematics, quite straightforward in the sense that one
measures $\alpha$ directly. On the other hand, measurements using Oklo and
meteorite data are indirect: what one measures directly here
is some combination of various couplings, and using them to obtain
constraints on $\alpha$ requires either assuming that other couplings don't
vary (which is almost certainly unrealistic) or assuming some (necessarily
model-dependent) relations between them. It is therefore important to check
how robust these constraints are to the specific assumptions being made
to obtain them

Of course, if both of these observational results survive further scrutiny,
then our toy model cannot be correct. We
emphasize again that any realistic model will reduce to a model in this
class for some period of time close to today, so that would indicate
that our linearized approximation will break down very close to today,
arguably much earlier than one would have thought. Note also that this
class of models, with a linearized behaviour for the scalar field,
are arguably the \textit{simplest} possible
models for a varying $\alpha$. Certainly models where $\alpha$ has a
linear dependence on redshift or on cosmic time (which have been explicitly
or implicitly assumed by a number of authors) are much more un-natural, and
it's hard to see how such could be obtained from a sensible particle
physics theory an a way that would be consistent with other observational and
experimental constraints.

\section{Conclusions}

We have studied the simplest class of Bekenstein-type, varying $\alpha$ models,
and compared them to existing observational constraints. These are models
in which the two available free functions (the potential and the gauge kinetic
function) are Taylor-expanded around present-day values, with terms
kept only up to linear order. Despite their apparent simplicity, they
are interesting to the extent that any realistic model of
this type should reduce to a model in this class for a certain time interval
around the present day. Nevertheless, their simplicity means that very
specific predictions ensue, that can be compared with existing data.
We have shown that no such model 
is consistent with all the existing observational results. Hence either some of
these observations are dominated by unknown systematics or our linearity
assumption breaks down on a timescale significantly smaller than a Hubble
time.

Given that a scalar field that produces a varying fine-structure constant
can also make a significant contribution towards the dark energy of the
universe, it's interesting to speculate on the possible relation between
the above observation and hints for a time-varying equation of state of dark
energy.Indeed in the latter context it has been argued that something
analogous seems to happen: observational  data seem to 
disfavour not only a constant equation of state, but even a mildly
varying one, say with a linear dependence in redshift \cite{Bassett,Beca,Alam}.
It is unclear if the two things are somehow related, but it has been said
that a coincidence is always worth noticing---one can always discard it
later if it turns out to be just a coincidence.

%\begin{acknowledgments} 
%\end{acknowledgments} 

\bibliography{alpha} 

\begin{thebibliography}{36}
\expandafter\ifx\csname natexlab\endcsname\relax\def\natexlab#1{#1}\fi
\expandafter\ifx\csname bibnamefont\endcsname\relax
  \def\bibnamefont#1{#1}\fi
\expandafter\ifx\csname bibfnamefont\endcsname\relax
  \def\bibfnamefont#1{#1}\fi
\expandafter\ifx\csname citenamefont\endcsname\relax
  \def\citenamefont#1{#1}\fi
\expandafter\ifx\csname url\endcsname\relax
  \def\url#1{\texttt{#1}}\fi
\expandafter\ifx\csname urlprefix\endcsname\relax\def\urlprefix{URL }\fi
\providecommand{\bibinfo}[2]{#2}
\providecommand{\eprint}[2][]{\url{#2}}

\bibitem[{\citenamefont{Polchinski}(1998)}]{Polchinski}
\bibinfo{author}{\bibfnamefont{J.}~\bibnamefont{Polchinski}},
  \emph{\bibinfo{title}{String Theory (2 vols)}} (\bibinfo{year}{1998}),
  \bibinfo{note}{{ }Cambridge, U.K.: University Press}.

\bibitem[{\citenamefont{Martins}(2002)}]{Essay}
\bibinfo{author}{\bibfnamefont{C.~J. A.~P.} \bibnamefont{Martins}},
  \bibinfo{journal}{Phil. Trans. Roy. Soc. Lond.}
  \textbf{\bibinfo{volume}{A360}}, \bibinfo{pages}{2681}
  (\bibinfo{year}{2002}), \eprint{astro-ph/0205504}.

\bibitem[{\citenamefont{Uzan}(2003)}]{Uzan}
\bibinfo{author}{\bibfnamefont{J.-P.} \bibnamefont{Uzan}},
  \bibinfo{journal}{Rev. Mod. Phys.} \textbf{\bibinfo{volume}{75}},
  \bibinfo{pages}{403} (\bibinfo{year}{2003}), \eprint{hep-ph/0205340}.

\bibitem[{\citenamefont{Martins}(2003)}]{VFC}
\bibinfo{author}{\bibfnamefont{C.~J. A.~P.} \bibnamefont{Martins}},
  \emph{\bibinfo{title}{The Cosmology of Extra Dimensions and Varying
  Fundamental Constants: A JENAM2002 Workshop}} (\bibinfo{year}{2003}),
  \bibinfo{note}{{ }Dordrecht, The Netherlands: Kluwer Academic Publishers}.

\bibitem[{\citenamefont{Will}(2001)}]{Will}
\bibinfo{author}{\bibfnamefont{C.~M.} \bibnamefont{Will}},
  \bibinfo{journal}{Living Rev. Rel.} \textbf{\bibinfo{volume}{4}},
  \bibinfo{pages}{4} (\bibinfo{year}{2001}), \eprint{gr-qc/0103036}.

\bibitem[{\citenamefont{Damour}(2003)}]{Damour2}
\bibinfo{author}{\bibfnamefont{T.}~\bibnamefont{Damour}}
  (\bibinfo{year}{2003}), \eprint{gr-qc/0306023}.

\bibitem[{\citenamefont{Tonry et~al.}(2003)}]{Tonry}
\bibinfo{author}{\bibfnamefont{J.~L.} \bibnamefont{Tonry}}
  \bibnamefont{et~al.}, \bibinfo{journal}{Astrophys. J.}
  \textbf{\bibinfo{volume}{594}}, \bibinfo{pages}{1} (\bibinfo{year}{2003}),
  \eprint{astro-ph/0305008}.

\bibitem[{\citenamefont{Bennett et~al.}(2003)}]{Bennett}
\bibinfo{author}{\bibfnamefont{C.~L.} \bibnamefont{Bennett}}
  \bibnamefont{et~al.} (\bibinfo{year}{2003}), \eprint{astro-ph/0302207}.

\bibitem[{\citenamefont{Bernardeau}(2003)}]{Bernardeau}
\bibinfo{author}{\bibfnamefont{F.}~\bibnamefont{Bernardeau}},
  \bibinfo{journal}{Rept. Prog. Phys.} \textbf{\bibinfo{volume}{66}},
  \bibinfo{pages}{691} (\bibinfo{year}{2003}).

\bibitem[{\citenamefont{Webb et~al.}(2001)}]{Webb}
\bibinfo{author}{\bibfnamefont{J.~K.} \bibnamefont{Webb}} \bibnamefont{et~al.},
  \bibinfo{journal}{Phys. Rev. Lett.} \textbf{\bibinfo{volume}{87}},
  \bibinfo{pages}{091301} (\bibinfo{year}{2001}), \eprint{astro-ph/0012539}.

\bibitem[{\citenamefont{Murphy et~al.}(2003)\citenamefont{Murphy, Webb, and
  Flambaum}}]{Murphy}
\bibinfo{author}{\bibfnamefont{M.~T.} \bibnamefont{Murphy}},
  \bibinfo{author}{\bibfnamefont{J.~K.} \bibnamefont{Webb}}, \bibnamefont{and}
  \bibinfo{author}{\bibfnamefont{V.~V.} \bibnamefont{Flambaum}},
  \bibinfo{journal}{Mon. Not. Roy. Astron. Soc.}
  \textbf{\bibinfo{volume}{345}}, \bibinfo{pages}{609} (\bibinfo{year}{2003}),
  \eprint{astro-ph/0306483}.

\bibitem[{\citenamefont{Chand et~al.}(2004)\citenamefont{Chand, Srianand,
  Petitjean, and Aracil}}]{Petit}
\bibinfo{author}{\bibfnamefont{H.}~\bibnamefont{Chand}},
  \bibinfo{author}{\bibfnamefont{R.}~\bibnamefont{Srianand}},
  \bibinfo{author}{\bibfnamefont{P.}~\bibnamefont{Petitjean}},
  \bibnamefont{and} \bibinfo{author}{\bibfnamefont{B.}~\bibnamefont{Aracil}}
  (\bibinfo{year}{2004}), \eprint{astro-ph/0401094}.

\bibitem[{\citenamefont{Fujii}(2003)}]{Fujii}
\bibinfo{author}{\bibfnamefont{Y.}~\bibnamefont{Fujii}},
  \bibinfo{journal}{Astrophys. Space Sci.} \textbf{\bibinfo{volume}{283}},
  \bibinfo{pages}{559} (\bibinfo{year}{2003}), \eprint{gr-qc/0212017}.

\bibitem[{\citenamefont{Ivanchik et~al.}(2003)\citenamefont{Ivanchik,
  Petitjean, Rodriguez, and Varshalovich}}]{Ivanchik}
\bibinfo{author}{\bibfnamefont{A.}~\bibnamefont{Ivanchik}},
  \bibinfo{author}{\bibfnamefont{P.}~\bibnamefont{Petitjean}},
  \bibinfo{author}{\bibfnamefont{E.}~\bibnamefont{Rodriguez}},
  \bibnamefont{and}
  \bibinfo{author}{\bibfnamefont{D.}~\bibnamefont{Varshalovich}},
  \bibinfo{journal}{Astrophys. Space Sci.} \textbf{\bibinfo{volume}{283}},
  \bibinfo{pages}{583} (\bibinfo{year}{2003}), \eprint{astro-ph/0210299}.

\bibitem[{\citenamefont{Marion et~al.}(2003)}]{Marion}
\bibinfo{author}{\bibfnamefont{H.}~\bibnamefont{Marion}} \bibnamefont{et~al.},
  \bibinfo{journal}{Phys. Rev. Lett.} \textbf{\bibinfo{volume}{90}},
  \bibinfo{pages}{150801} (\bibinfo{year}{2003}), \eprint{physics/0212112}.

\bibitem[{\citenamefont{Olive et~al.}(2004)}]{Olive}
\bibinfo{author}{\bibfnamefont{K.~A.} \bibnamefont{Olive}}
  \bibnamefont{et~al.}, \bibinfo{journal}{Phys. Rev.}
  \textbf{\bibinfo{volume}{D69}}, \bibinfo{pages}{027701}
  (\bibinfo{year}{2004}), \eprint{astro-ph/0309252}.

\bibitem[{\citenamefont{Avelino et~al.}(2001)}]{Avelino}
\bibinfo{author}{\bibfnamefont{P.~P.} \bibnamefont{Avelino}}
  \bibnamefont{et~al.}, \bibinfo{journal}{Phys. Rev.}
  \textbf{\bibinfo{volume}{D64}}, \bibinfo{pages}{103505}
  (\bibinfo{year}{2001}), \eprint{astro-ph/0102144}.

\bibitem[{\citenamefont{Martins et~al.}(2002)}]{Martins}
\bibinfo{author}{\bibfnamefont{C.~J. A.~P.} \bibnamefont{Martins}}
  \bibnamefont{et~al.}, \bibinfo{journal}{Phys. Rev.}
  \textbf{\bibinfo{volume}{D66}}, \bibinfo{pages}{023505}
  (\bibinfo{year}{2002}), \eprint{astro-ph/0203149}.

\bibitem[{\citenamefont{Rocha et~al.}(2003{\natexlab{a}})}]{Rocha}
\bibinfo{author}{\bibfnamefont{G.}~\bibnamefont{Rocha}} \bibnamefont{et~al.},
  \bibinfo{journal}{New Astron. Rev.} \textbf{\bibinfo{volume}{47}},
  \bibinfo{pages}{863} (\bibinfo{year}{2003}{\natexlab{a}}),
  \eprint{astro-ph/0309205}.

\bibitem[{\citenamefont{Martins et~al.}(2003)}]{wmap1}
\bibinfo{author}{\bibfnamefont{C.~J. A.~P.} \bibnamefont{Martins}}
  \bibnamefont{et~al.} (\bibinfo{year}{2003}), \eprint{astro-ph/0302295}.

\bibitem[{\citenamefont{Rocha et~al.}(2003{\natexlab{b}})}]{wmap2}
\bibinfo{author}{\bibfnamefont{G.}~\bibnamefont{Rocha}} \bibnamefont{et~al.}
  (\bibinfo{year}{2003}{\natexlab{b}}), \eprint{astro-ph/0309211}.

\bibitem[{\citenamefont{Bekenstein}(1982)}]{Bekenstein}
\bibinfo{author}{\bibfnamefont{J.~D.} \bibnamefont{Bekenstein}},
  \bibinfo{journal}{Phys. Rev.} \textbf{\bibinfo{volume}{D25}},
  \bibinfo{pages}{1527} (\bibinfo{year}{1982}).

\bibitem[{\citenamefont{Sandvik et~al.}(2002)\citenamefont{Sandvik, Barrow, and
  Magueijo}}]{Sandvik}
\bibinfo{author}{\bibfnamefont{H.~B.} \bibnamefont{Sandvik}},
  \bibinfo{author}{\bibfnamefont{J.~D.} \bibnamefont{Barrow}},
  \bibnamefont{and} \bibinfo{author}{\bibfnamefont{J.}~\bibnamefont{Magueijo}},
  \bibinfo{journal}{Phys. Rev. Lett.} \textbf{\bibinfo{volume}{88}},
  \bibinfo{pages}{031302} (\bibinfo{year}{2002}), \eprint{astro-ph/0107512}.

\bibitem[{\citenamefont{Olive and Pospelov}(2002)}]{OlivePos}
\bibinfo{author}{\bibfnamefont{K.~A.} \bibnamefont{Olive}} \bibnamefont{and}
  \bibinfo{author}{\bibfnamefont{M.}~\bibnamefont{Pospelov}},
  \bibinfo{journal}{Phys. Rev.} \textbf{\bibinfo{volume}{D65}},
  \bibinfo{pages}{085044} (\bibinfo{year}{2002}), \eprint{hep-ph/0110377}.

\bibitem[{\citenamefont{Anchordoqui and Goldberg}(2003)}]{Anchordoqui}
\bibinfo{author}{\bibfnamefont{L.}~\bibnamefont{Anchordoqui}} \bibnamefont{and}
  \bibinfo{author}{\bibfnamefont{H.}~\bibnamefont{Goldberg}},
  \bibinfo{journal}{Phys. Rev.} \textbf{\bibinfo{volume}{D68}},
  \bibinfo{pages}{083513} (\bibinfo{year}{2003}), \eprint{hep-ph/0306084}.

\bibitem[{\citenamefont{Parkinson et~al.}(2004)\citenamefont{Parkinson,
  Bassett, and Barrow}}]{Parkinson}
\bibinfo{author}{\bibfnamefont{D.}~\bibnamefont{Parkinson}},
  \bibinfo{author}{\bibfnamefont{B.~A.} \bibnamefont{Bassett}},
  \bibnamefont{and} \bibinfo{author}{\bibfnamefont{J.~D.}
  \bibnamefont{Barrow}}, \bibinfo{journal}{Phys. Lett.}
  \textbf{\bibinfo{volume}{B578}}, \bibinfo{pages}{235} (\bibinfo{year}{2004}),
  \eprint{astro-ph/0307227}.

\bibitem[{\citenamefont{Copeland et~al.}(2004)\citenamefont{Copeland, Nunes,
  and Pospelov}}]{Copeland}
\bibinfo{author}{\bibfnamefont{E.~J.} \bibnamefont{Copeland}},
  \bibinfo{author}{\bibfnamefont{N.~J.} \bibnamefont{Nunes}}, \bibnamefont{and}
  \bibinfo{author}{\bibfnamefont{M.}~\bibnamefont{Pospelov}},
  \bibinfo{journal}{Phys. Rev.} \textbf{\bibinfo{volume}{D69}},
  \bibinfo{pages}{023501} (\bibinfo{year}{2004}), \eprint{hep-ph/0307299}.

\bibitem[{\citenamefont{Nunes and Lidsey}(2003)}]{Nunes}
\bibinfo{author}{\bibfnamefont{N.~J.} \bibnamefont{Nunes}} \bibnamefont{and}
  \bibinfo{author}{\bibfnamefont{J.~E.} \bibnamefont{Lidsey}}
  (\bibinfo{year}{2003}), \eprint{astro-ph/0310882}.

\bibitem[{\citenamefont{Damour and Nordtvedt}(1993)}]{Damour}
\bibinfo{author}{\bibfnamefont{T.}~\bibnamefont{Damour}} \bibnamefont{and}
  \bibinfo{author}{\bibfnamefont{K.}~\bibnamefont{Nordtvedt}},
  \bibinfo{journal}{Phys. Rev.} \textbf{\bibinfo{volume}{D48}},
  \bibinfo{pages}{3436} (\bibinfo{year}{1993}).

\bibitem[{\citenamefont{Santiago et~al.}(1998)\citenamefont{Santiago, Kalligas,
  and Wagoner}}]{Santiago}
\bibinfo{author}{\bibfnamefont{D.~I.} \bibnamefont{Santiago}},
  \bibinfo{author}{\bibfnamefont{D.}~\bibnamefont{Kalligas}}, \bibnamefont{and}
  \bibinfo{author}{\bibfnamefont{R.~V.} \bibnamefont{Wagoner}},
  \bibinfo{journal}{Phys. Rev.} \textbf{\bibinfo{volume}{D58}},
  \bibinfo{pages}{124005} (\bibinfo{year}{1998}), \eprint{gr-qc/9805044}.

\bibitem[{\citenamefont{Hannestad and Mortsell}(2002)}]{Steen}
\bibinfo{author}{\bibfnamefont{S.}~\bibnamefont{Hannestad}} \bibnamefont{and}
  \bibinfo{author}{\bibfnamefont{E.}~\bibnamefont{Mortsell}},
  \bibinfo{journal}{Phys. Rev.} \textbf{\bibinfo{volume}{D66}},
  \bibinfo{pages}{063508} (\bibinfo{year}{2002}), \eprint{astro-ph/0205096}.

\bibitem[{\citenamefont{Melchiorri et~al.}(2003)\citenamefont{Melchiorri,
  Mersini, Odman, and Trodden}}]{Melchiorri}
\bibinfo{author}{\bibfnamefont{A.}~\bibnamefont{Melchiorri}},
  \bibinfo{author}{\bibfnamefont{L.}~\bibnamefont{Mersini}},
  \bibinfo{author}{\bibfnamefont{C.~J.} \bibnamefont{Odman}}, \bibnamefont{and}
  \bibinfo{author}{\bibfnamefont{M.}~\bibnamefont{Trodden}},
  \bibinfo{journal}{Phys. Rev.} \textbf{\bibinfo{volume}{D68}},
  \bibinfo{pages}{043509} (\bibinfo{year}{2003}), \eprint{astro-ph/0211522}.

\bibitem[{\citenamefont{Jimenez et~al.}(2003)\citenamefont{Jimenez, Verde,
  Treu, and Stern}}]{Jimenez}
\bibinfo{author}{\bibfnamefont{R.}~\bibnamefont{Jimenez}},
  \bibinfo{author}{\bibfnamefont{L.}~\bibnamefont{Verde}},
  \bibinfo{author}{\bibfnamefont{T.}~\bibnamefont{Treu}}, \bibnamefont{and}
  \bibinfo{author}{\bibfnamefont{D.}~\bibnamefont{Stern}},
  \bibinfo{journal}{Astrophys. J.} \textbf{\bibinfo{volume}{593}},
  \bibinfo{pages}{622} (\bibinfo{year}{2003}), \eprint{astro-ph/0302560}.

\bibitem[{\citenamefont{Bassett et~al.}(2003)\citenamefont{Bassett, Kunz,
  Parkinson, and Ungarelli}}]{Bassett}
\bibinfo{author}{\bibfnamefont{B.~A.} \bibnamefont{Bassett}},
  \bibinfo{author}{\bibfnamefont{M.}~\bibnamefont{Kunz}},
  \bibinfo{author}{\bibfnamefont{D.}~\bibnamefont{Parkinson}},
  \bibnamefont{and}
  \bibinfo{author}{\bibfnamefont{C.}~\bibnamefont{Ungarelli}},
  \bibinfo{journal}{Phys. Rev.} \textbf{\bibinfo{volume}{D68}},
  \bibinfo{pages}{043504} (\bibinfo{year}{2003}), \eprint{astro-ph/0211303}.

\bibitem[{\citenamefont{Avelino et~al.}(2003)\citenamefont{Avelino, Beca,
  de~Carvalho, Martins, and Copeland}}]{Beca}
\bibinfo{author}{\bibfnamefont{P.~P.} \bibnamefont{Avelino}},
  \bibinfo{author}{\bibfnamefont{L.~M.~G.} \bibnamefont{Beca}},
  \bibinfo{author}{\bibfnamefont{J.~P.~M.} \bibnamefont{de~Carvalho}},
  \bibinfo{author}{\bibfnamefont{C.~J. A.~P.} \bibnamefont{Martins}},
  \bibnamefont{and} \bibinfo{author}{\bibfnamefont{E.~J.}
  \bibnamefont{Copeland}} (\bibinfo{year}{2003}), \eprint{astro-ph/0306493}.

\bibitem[{\citenamefont{Alam et~al.}(2003)\citenamefont{Alam, Sahni, Saini, and
  Starobinsky}}]{Alam}
\bibinfo{author}{\bibfnamefont{U.}~\bibnamefont{Alam}},
  \bibinfo{author}{\bibfnamefont{V.}~\bibnamefont{Sahni}},
  \bibinfo{author}{\bibfnamefont{T.~D.} \bibnamefont{Saini}}, \bibnamefont{and}
  \bibinfo{author}{\bibfnamefont{A.~A.} \bibnamefont{Starobinsky}}
  (\bibinfo{year}{2003}), \eprint{astro-ph/0311364}.

\end{thebibliography}
\end{document}